\documentclass{article}
\usepackage{spconf,amsmath,graphicx}
\usepackage{bm} 							
\usepackage{hyperref} 					
\usepackage{url}							
\usepackage{subcaption} 				
\usepackage{amssymb}
\usepackage{mathtools} 					
\usepackage{multirow} 					
\usepackage{hyperref}
\usepackage{booktabs} 					

\newcommand{\x} {\bm{x}}
\newcommand{\y} {\bm{y}}
\newcommand{\h} {\bm{h}}
\newcommand{\z} {\bm{z}}
\newcommand{\e} {\mathbf{e}}
\newcommand{\vecl} {\mathbf{l}}

\newcommand{\vecc} {\bm{c}}

\newcommand{\veci} {\mathbf{i}}

\newcommand{\Enc} {\text{Enc}}
\newcommand{\Dec} {\text{Dec}}
\newcommand{\sep} {\textit{separate}}
\newcommand{\comb} {\textit{combine}}
\DeclareMathOperator*{\argmax}{arg\,max}
\DeclareMathOperator*{\Exp}{\mathbb{E}} 	

\newcommand{\vqw} {VQW2V}
\title{ANY-TO-ONE SEQUENCE-TO-SEQUENCE VOICE CONVERSION USING SELF-SUPERVISED DISCRETE SPEECH REPRESENTATIONS}

\name{Wen-Chin Huang, Yi-Chiao Wu, Tomoki Hayashi, Tomoki Toda}
\address{Nagoya University, Japan}

\begin{document}
\ninept
\maketitle

\begin{abstract}
We present a novel approach to any-to-one (A2O) voice conversion (VC) in a sequence-to-sequence (seq2seq) framework. A2O VC aims to convert any speaker, including those unseen during training, to a fixed target speaker. We utilize vq-wav2vec (VQW2V), a discretized self-supervised speech representation that was learned from massive unlabeled data, which is assumed to be speaker-independent and well corresponds to underlying linguistic contents. Given a training dataset of the target speaker, we extract VQW2V and acoustic features to estimate a seq2seq mapping function from the former to the latter. With the help of a pretraining method and a newly designed postprocessing technique, our model can be generalized to only 5 min of data, even outperforming the same model trained with parallel data.
\end{abstract}
\begin{keywords}
voice conversion, any-to-one voice conversion, self-supervised speech representation, vq-wav2vec, sequence-to-sequence modeling
\end{keywords}
\section{Introduction}
\label{sec:intro}

Voice conversion (VC) aims to convert a speech from a source to that of a target without changing the linguistic content \cite{VC}. In recent years, compared with conventional frame-wise VC methods, sequence-to-sequence (seq2seq) \cite{S2S}-based VC has been a promising approach in terms of conversion similarity \cite{ATT-S2S-VC, CONV-S2S-VC, S2S-iFLYTEK-VC, S2S-NP-VC, S2S-parrotron-VC, VTN}. Its capability to generate outputs of various lengths and capture long-term dependencies makes it a suitable choice to handle the suprasegmental characteristics of F0 and duration patterns, which are closely correlated with the speaker identity.

Despite promising results, most seq2seq VC systems can only model parallel one-to-one (O2O) VC, which is only capable of converting from a known source to a known target with the requirement of a parallel training dataset. Some attempts have been made to relax this constraint. For instance, a nonparallel training framework was proposed in \cite{S2S-NP-VC} where text labels were utilized to learn meaningful hidden representations, whereas a complex model and objectives required rigorous hyperparameter tuning. On the other hand, any-to-one (A2O) VC aims to convert from any unseen source to a known target, which is attractive owing to its flexibility. In \cite{S2S-parrotron-VC}, a framework for A2O VC was proposed, where a text-to-speech (TTS) system was used to generate 30-k-h synthetic parallel data of the target speaker. However, such a huge amount of data is impractical to collect.

In recent years, self-supervised speech representation learning has been an active field of study, which aims to clarify compact, high-level representations that can benefit potential downstream tasks. When either an autoencoding objective \cite{wnae, APC, MT-APC, mockingjay} or a contrastive loss \cite{CPC, wav2vec, vq-wav2vec} is employed, requirements on manual labels are no longer needed, which provides benefits from massive, unlabeled speech data. However, most of them were shown effective only on tasks such as speech recognition or speaker identification, and there has been no evidence on their effectiveness for VC.

In this work, we propose a novel approach that utilizes vq-wav2vec (\vqw), a discretized self-supervised speech representation \cite{vq-wav2vec} for A2O seq2seq VC.
The discretization operation in the \vqw\ model allows us to eliminate speaker information and meanwhile represent the speech signal by a sequence of indices of the corresponding codewords, which are believed to be the underlying spoken contents.
Then, a seq2seq model can be trained to estimate the mapping function from the \vqw\ to the acoustic features of the target speaker.
Our approach can then be formulated as training a target-speaker-dependent TTS model that can take as input the discrete \vqw\ from any source speech to generate speech as spoken by the target speaker.
To tackle the limited size of training data, in addition to a pretraining scheme \cite{VTN}, we propose to further use a postprocessing technique that takes advantage of the multiple quantization group strategy, which was originally used to solve the mode collapse problem in \cite{vq-wav2vec}. As we demonstrate in later sections, these methods are extremely effective against limited training data.
Our contributions in this work are as follows:
\begin{itemize}
	\item We apply \vqw, a discretized self-supervised speech representation to VC. 
	\item By employing the pretraining technique and the postprocessing technique for \vqw\ indices, we are the first to successfully train a seq2seq A2O VC model with only 5 min of \textit{target} training data.
	\item Our proposed A2O VC system does not use any parallel data and can perform comparably well to an O2O VC system \cite{VTN}.
\end{itemize}

\section{Background}
\label{sec:background}

\subsection{Sequence-to sequence acoustic modeling}
\label{ssec:seq2seq-am}

Seq2seq models learn mapping between a source feature sequence $\mathbf{X}=\x_{1:n}=(\x_1, \cdots, \x_n)$ and a target feature sequence $\mathbf{Y}=\y_{1:m}=(\y_1, \cdots, \y_m)$, which are often of different lengths, i.e., $n \neq m$. Most seq2seq models have an encoder\textendash decoder structure \cite{S2S}. The encoder ($\Enc$) first maps the input acoustic feature sequence $\x_{1:n}$ into a sequence of hidden representations $\mathbf{H}=\h_{1:n}= \Enc(\x_{1:n})$. The decoder ($\Dec$) is autoregressive, which means that when decoding the current output $\y_t$, in addition to the encoder output, i.e., the hidden representations $\h_{1:n}$, the previously generated features $\y_{1:t-1}$ are also considered, i.e., $\y_t = \Dec(\h_{1:n}, \y_{1:t-1})$. The training objective includes an L1 loss, in combination with a weighted binary cross-entropy loss on the stop token prediction. The entire network is composed of neural networks and optimized via backpropagation.


The same seq2seq framework can be used to model both discrete and continuous inputs by using different front-end modules. When employing the model for TTS, a simple embedding lookup table is used. For parallel acoustic feature mappings (e.g., parallel O2O VC) following \cite{transformer-asr}, the input acoustic feature sequence is downsampled in both time and frequency axes by a fraction of 4 using two convolutional layers with stride $2\times 2$, followed by a linear projection.

\subsection{\vqw}
\label{ssec:vqw}

\vqw\ \cite{vq-wav2vec} aims to learn contextualized speech representations by trying to predict the future \cite{CPC} with a vector quantization module that adds discreteness into the representations. The network is composed of an encoder $f: \mathcal{X} \to \mathcal{Z}$ for feature extraction, a quantizer $q: \mathcal{Z} \to \mathcal{\hat{Z}}$ to facilitate discreteness, and an aggregator $g: \mathcal{\hat{Z}} \to \mathcal{C}$ for aggregation. First, 30 ms segments of raw speech signals are mapped to a local feature $\z$ with a 10 ms stride using the encoder $f$. Then, the quantizer $q$ turns the local features $\z$ into discrete indices, which are used to choose reconstruction vectors $\hat{\z}$ of the original representations. The aggregator finally takes the quantized $\hat{\z}$ to generate a contextualized representation $\vecc$.

\begin{figure}[t]
  \centering
  \includegraphics[width=0.45\textwidth]{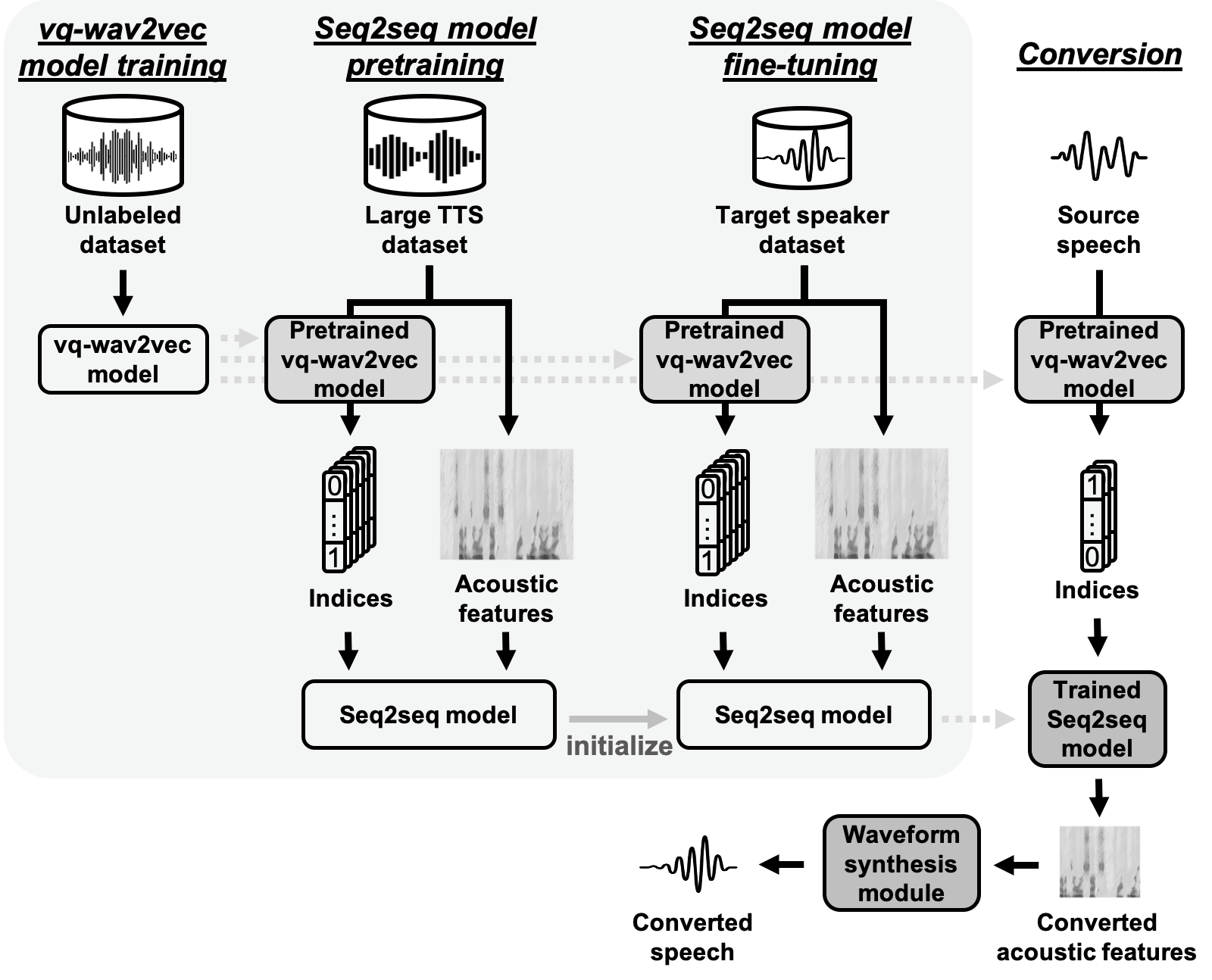}
  \centering
  \captionof{figure}{Illustration of the proposed method.}
  \label{fig:proposed}
  \vspace*{-3mm}
\end{figure}

\subsubsection{Optimization}
\label{sssec:vqw-optimization}

The\vqw\ model is optimized via a context prediction task, which is also known as contrastive loss. Given a contextualized representation $\vecc_i$, the model is trained to distinguish $\z_{i+k}$, a positive sample that is $k$ steps in the future, from negative samples $\tilde{\z}$ drawn from a uniform distribution $p_n$ over the entire audio sequence by minimizing the contrastive loss for steps $k=1,\dots,K$:
\begin{equation}
	  \mathcal{L}_k = - \sum_{i=1}^{T - k} \Big(
  \log \sigma(\z_{i+k}^\top \mathbf{W}_k \vecc_i) \\
  + \lambda \Exp_{\mathclap{\tilde{\z} \sim p_n}}\ [ \log \sigma(-\tilde{\z}^\top  \mathbf{W}_k \vecc_i) ]\ \Big),
  \label{eq:contrastive-loss}
\end{equation}
where $T$ is the sequence length, $\mathbf{W}_k$ is a stepwise learnable affine matrix, $\sigma(x) = 1/(1+\exp(-x))$ is the sigmoid function, and $\sigma(\z_{i+k}^\top h_k(\vecc_i))$ is the probability of $\z_{i+k}$ being the true sample. The loss in~(\ref{eq:contrastive-loss}) is summed over different step sizes, resulting in a final loss $\mathcal{L} = \sum_{k=1}^K \mathcal{L}_k$.

\subsubsection{Quantization with a partitioning technique}
\label{sssec:quantization}

The quantization module $q$ replaces an input feature vector $\z$ by $\hat{\z} = \e_i$ from a fixed size codebook $\e \in \mathbb{R}^{V \times d}$ containing $V$ codewords with dimension $d$. As in \cite{vq-wav2vec}, the Gumbel-softmax was employed, which is a differentiable approximation of the argmax for computing one-hot representations \cite{gumbel-softmax}. Given a feature vector $\z$, a logits vector $\vecl \in \mathbb{R}^V$ is calculated for the Gumbel-softmax. During training, the probabilities for choosing the $j$-th codeword are
\begin{equation}
	p_j = \frac{\exp(l_j+v_j)/\tau}{\sum_{k=1}^V \exp(l_k+v_k)/\tau},
\end{equation}
where $v = -\log(-\log(u))$ and $u$ are uniform samples from $U(0,1)$. During the forward pass, the largest index is simply selected, i.e., $i = \argmax_j p_j$, and in the backward pass, the true gradient of the Gumbel-softmax outputs is used.

Replacing the feature vector $\z$ with a single codebook entry $\e_i$ can make the model prone to mode collapse, where only some of the codewords are activated. In \cite{vq-wav2vec}, a partitioning technique was proposed as a remedy. Each feature vector $\z \in \mathbb{R}^d$ is first partitioned into $G$ groups: $\z' \in \mathbb{R}^{d/G}$. Then, for each $\z'$, the quantization process described above is applied. That is, the full feature vector can be represented by a sequence of indices $\veci \in [V]^G$, where each element $\veci_j$ corresponds to a codeword. Following \cite{vq-wav2vec}, a single codebook of size $\e \in \mathbb{R}^{V \times (d/G)}$ is shared across different groups.

\section{Proposed Method}
\label{sec:method}

\subsection{Overview}

Figure~\ref{fig:proposed} illustrates our proposed method. The core functions are a \vqw\ model that encodes the raw input speech into discretized features, which are represented as indices, and a target-speaker-dependent seq2seq model that maps them to acoustic features for speech generation.

\begin{figure}[t]
  \centering
  \includegraphics[width=0.3\textwidth]{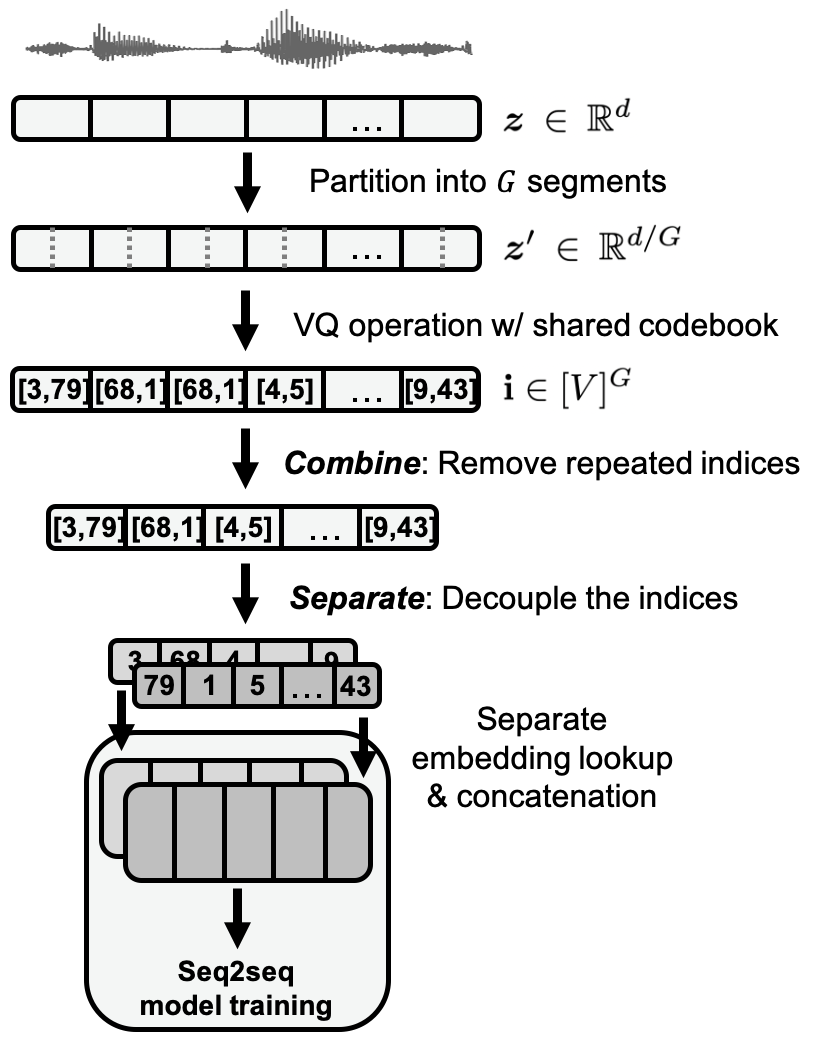}
  \centering
  \captionof{figure}{Illustration of the postprocessing procedure for \vqw\ indices. In this example, the codebook size is set to 320 and the group size is set to 2.}
  \label{fig:vq-indices-postprocessing}
  \vspace*{-6mm}
\end{figure}

\begin{table*}[t]
	\centering
	\captionsetup{justification=centering}
	\caption{Results of objective evaluation using the validation set of the baseline VTN and variants of the proposed method trained on different sizes of training data.}
	\centering
	\begin{tabular}{ c c c c c c c c c c }
		\toprule
		 & & \multicolumn{2}{c}{Postprocessing} & \multicolumn{3}{c}{932 training utterances} & \multicolumn{3}{c}{80 training utterances} \\
		\cmidrule(lr){3-4} \cmidrule(lr){5-7} \cmidrule(lr){8-10}
		Speaker & Description & Combine & Separate & MCD & CER & WER & MCD & CER & WER \\
		\midrule
		\multirow{4}{*}[-1pt]{clb-slt} & \multirow{3}{*}{Proposed} & - & - & 6.57 & 2.9 & 6.4 & 9.52 & 82.2 & 106.5 \\
		& & - 		   & \checkmark & 6.60 & 2.6 & 6.0 & 6.88 & 6.7 & 10.0 \\
		& & \checkmark & \checkmark & \textbf{6.56} & \textbf{3.0} & \textbf{7.3} & \textbf{6.62} & \textbf{3.1} & \textbf{7.5} \\
		\cmidrule(lr){2-10} 
		& VTN & - & - & 6.02 & 5.5 & 9.1 & 6.66 & 10.4 & 14.7   \\
		\midrule
		\multirow{4}{*}[-1pt]{bdl-slt} & \multirow{3}{*}{Proposed} & - & - & 6.68 & 3.8 & 7.9 & 9.53 & 71.6 & 92.3 \\
		& & - & \checkmark & 6.82 & 4.3 & 8.9 & 7.04 & 8.0 & 12.6 \\
		& & \checkmark & \checkmark & \textbf{6.77} & \textbf{5.6} & \textbf{11.3} & \textbf{6.89} & \textbf{5.1} & \textbf{11.3} \\
		\cmidrule(lr){2-10} 
		& VTN & - & - & 6.33 & 5.0 & 7.6 & 7.07 & 9.7 & 13.6   \\
		\bottomrule
	\end{tabular}
	\label{tab:obj-eval}
	\vspace*{-3mm}
\end{table*}

\subsection{Postprocessing of \vqw\ indices}
\label{ssec:vq-indices-postprocessing}

The \vqw\ model is first trained on a large unlabeled corpus as described in Section~\ref{ssec:vqw}. The pretrained model is then used as a feature extractor in the seq2seq model training phases and the conversion phase. Our approach utilizes the quantization characteristic of \vqw\ and forms a TTS task; thus, only the discrete indices $\veci$ are needed and the aggregator $g$ is discarded.

The partitioning technique described in Section~\ref{sssec:quantization} can exponentially increase the codebook size. For example, with $V=320$ and $G=2$, we obtain a vocabulary of size $V^G \approx 102k$, which can greatly increase the difficulty of training the seq2seq model because some entries in the embedding lookup table (Section~\ref{ssec:seq2seq-am}) might not be fully updated. To mitigate this problem, we propose a postprocessing strategy as depicted in Figure~\ref{fig:vq-indices-postprocessing}. Specifically, the postprocessing consists of two steps:
\begin{itemize}
	\item \textit{Combine}: The discrete indices $\veci$ are of frame-level resolution; thus, adjacent frames with similar characteristics may be quantized to the same index in the codebook, resulting in many repeated indices. Thus, as a first step, we combine the repeated indices, which can reduce the length by about $20\%$ on average.
	\item \textit{Separate}: Since the codebook is shared among groups, we assume that it resembles a hierarchical structure, similar to the relationship between prefixes and postfixes of English words or consonants and vowels of phonemes. On the basis of this assumption, we separate the indices of different groups and used them to look up different embedding tables in the seq2seq model. The resulting embeddings are then concatenated and sent into further layers. As a result, the total size of the embedding lookup table is reduced from $V^G$ to $V\cdot G$. This is particularly important when only a small target training set is available, as we will show in our experiments.
\end{itemize}

\subsection{Seq2seq model training}

The seq2seq model maps the \vqw\ discrete indices extracted from the speech to the acoustic features for speech waveform synthesis. In VC, a practical setting only allows access to around 5 min of speech of the target speaker, which is too limited for seq2seq model training. To solve this problem, we use a pretraining\textendash finetuning scheme to improve performance. Following \cite{VTN}, we first pretrain the seq2seq model on a large-scale TTS corpus, followed by fine-tuning on the target speaker dataset. 

\subsection{Conversion}

During conversion, given a source speech, the \vqw\ discrete indices are first extracted and postprocessed as described in Section~\ref{ssec:vq-indices-postprocessing}. Then, the indices are consumed by the seq2seq model to generate the converted acoustic features. Finally, a vocoder synthesizes the converted waveform from the converted acoustic features.

\vspace*{-2mm}
\section{Experimental settings}
\label{sec:experimental-settings}

\subsection{Datasets}

We evaluated our proposed method on the CMU ARCTIC database \cite{CMU-Arctic}, which contains parallel recordings of professional US English speakers sampled at 16 kHz.
A male speaker (\textit{rms}) and a female speaker (\textit{slt}) were chosen as the targets, and one seq2seq model was trained for each of them. Either a maximum of roughly 1-hr-long or 5-min-long utterances were used as training data.
During conversion, we used a male speaker (\textit{bdl}) and a female speaker (\textit{clb}) as the unseen source speakers. The validation and test sets each contained 100 utterances.

As for pretraining, the \vqw\ model used the full 960 hr Librispeech dataset \cite{librispeech}.
For the TTS pretraining data, we chose a US female English speaker (\textit{judy bieber}) from the M-AILABS speech dataset \cite{M-AILABS}. The dataset has a 16 kHz sampling rate and contains 15,200 audiobook recordings, which are roughly 32 hr long.

\subsection{Implementation}

For \vqw, we used the publicly available\footnote{\url{https://github.com/pytorch/fairseq}} pretrained model provided by fairseq \cite{fairseq}, which has eight convolutional layers with 512 channels each, and a total stride of 160. The indices have $G=2$ groups with $V=320$ codewords per group and generates $\approx$13.5k unique index combinations out of the 102k possible codewords. For the acoustic features, 80-dimensional mel filterbanks with 1024 FFT points and a 256 point frame shift (16 ms) were used.

For the seq2seq model, the implementation was carried out on the open-source ESPnet toolkit \cite{espnet-tts, espnet}. Our model is based on the Transformer-TTS architecture \cite{transformer-tts} as described in Section~\ref{ssec:seq2seq-am}. The model was pretrained with a batch size of 60 and fine-tuned with a batch size of 16. The detailed model and training configuration can be found online\footnote{\url{https://gist.github.com/unilight/a48f99cf6a47c0b4e5b96fe1d6e59397}}.
As for the baseline O2O VC model, we considered the Voice Transformer Network (VTN) with TTS pretraining \cite{VTN, VTN-TASLP} and followed the official implementation\footnote{\url{https://github.com/espnet/espnet/tree/master/egs/arctic/vc1}}.

For the waveform synthesis module, following \cite{VTN}, we used the Parallel WaveGAN (PWG) neural vocoder \cite{parallel-wavegan}, which enables parallel and faster than real-time waveform generation. We followed  an open-source implementation.

\subsection{Evaluation metrics and protocols}

We carried out two types of objective evaluation between the converted speech and the ground truth: the mel cepstrum distortion (MCD), a commonly used measure of spectral distortion in VC, and the character error rate (CER) as well as the word error rate (WER), which indicates intelligibility. The ASR engine was Transformer-based \cite{transformer-asr} and trained on LibriSpeech.

Systemwise subjective tests on naturalness and conversion similarity were also conducted to evaluate the perceptual performance.
For naturalness, participants were asked to evaluate the naturalness of the speech by the mean opinion score (MOS) test on a five-point scale.
For conversion similarity, each listener was presented a natural target speech and a converted speech, and asked to judge whether they were produced by the same speaker on a four-point scale.
For each system, 20 and 15 random utterances were chosen for the naturalness and similarity tests, respectively. All subjective evaluations were performed using the open-source toolkit \cite{p808-open-source} that implements the ITU-T Recommendation P.808 \cite{p808} for subjective speech quality assessment in the crowd using the Amazon Mechanical Turk (Mturk) and screens the obtained data for unreliable ratings. We recruited more than fifty listeners. Audio samples are available online\footnote{\url{https://unilight.github.io/Publication-Demos/publications/vq-wav2vec-vc/index.html}}.


\begin{table}[t]
	\centering
	\caption{Results of subjective evaluation using the test set with 95\% confidence intervals of the vocoder analysis\textendash synthesis, the baseline VTN and the proposed method. The numbers in parentheses indicate the number of training utterances.}
	\centering
	\begin{tabular}{ l c c }
		\toprule
		Description & Naturalness & Similarity \\
		\midrule
		Analysis\textendash synthesis & 3.78 $\pm$ 0.16 & - \\
		\midrule
		VTN (932) & 3.93 $\pm$ 0.14 & 74\% $\pm$ 4\% \\
		VTN (80) & 3.33 $\pm$ 0.18 & 58\% $\pm$ 5\% \\
		Proposed (932) & \textbf{3.71 $\pm$ 0.17} & \textbf{76\% $\pm$ 4\%} \\
		Proposed (80) & \textbf{3.76 $\pm$ 0.14} & \textbf{63\% $\pm$ 5\%} \\
		\bottomrule
	\end{tabular}
	\label{tab:sub-eval}
	\vspace*{-3mm}
\end{table}

\section{Experimental Evaluation Results}
\label{sec:results}

\subsection{Effectiveness of postprocessing of \vqw\ indices }
\label{ssec:effectiveness-tts-pt}

To evaluate the effectiveness of the postprocessing technique for \vqw\ indices  we proposed in Section~\ref{fig:vq-indices-postprocessing}, we conducted a systematic comparison across different sizes of training data. The results of objective evaluation are shown in Table~\ref{tab:obj-eval}. First, without any postprocessing, our method could already achieve satisfactory results with the full training set. However, as we reduced the training utterances from 932 (1 hr) to 80 (5 min), the performance dropped markedly. We suspect that the limited training data was insufficient to fully train all entries of the huge embedding table in the seq2seq model.

With the \sep\ operation, the performance was maintained and comparable to that with the full training set. In the case of 5 min training data, the performance reached a satisfactory level which is slightly inferior to the case of 1 hr of training data. This result clearly demonstrates the data efficiency of the \sep\ operation, which mainly comes from the benefit of exploiting the hierarchical structure of the \vqw\ indices to avoid the difficulty in training. Finally, by incorporating the \comb\ operation, we achieved further improvement especially on the limited training set. We consider that the \comb\ step eliminates the duration information of each text/phoneme from the input representation, forcing the model to implicitly capture the speaker-dependent duration pattern rather than the indices.

\vspace*{-2mm}
\subsection{Comparison with parallel O2O VC}

Next, we compared our method with VTN, the O2O parallel baseline. As shown in Table~\ref{tab:obj-eval}, our A2O method could outperform the O2O system with limited training data, especially in terms of intelligibility. The reason might be that in our method, the alignment learning in the seq2seq model was made much simpler since \vqw\ representations provide compact linguistic clues, whereas more interference existed when trying to align directly from source acoustic features. We also noticed that compared with VTN, our method does not suffer so much from training data reduction. That is to say, the marginal effect of increasing data for more than 5 min was relatively small. This demonstrates a superior data efficiency of our method over the VTN baseline.

Finally, Table~\ref{tab:sub-eval} shows the results of the subjective evaluation on the test set.
In terms of naturalness, the proposed method was slightly worse than VTN with 1 hr of training data, but was significantly better in the 5 min of training data, which is consistent with the findings in the objective evaluations.
As for similarity, the proposed method was comparable to VTN with the full training set and provided better performance when the training size was reduced to 5 min.
This result justifies the effectiveness of our method and also showed that it can greatly increase data efficiency without severe performance degradation.

\vspace*{-2mm}
\subsection{Credibility of subjective evaluation results}

Note that the analysis\textendash synthesis had a naturalness MOS slightly lower than that of the VTN trained with 932 utterances.
We calculated the p-value between the two systems and it was 0.22, showing that there was no statistically significant difference.
We suspect that the samples generated by VTN had more natural prosody patterns owing to the imperfect modeling of PWG.
It is also noticeable that our proposed system trained with 80 utterances had a slightly higher naturalness MOS than that of the proposed system trained with 932 utterances. We also calculated the p-value, and it was 0.66, showing no significant difference.
Taken together, statistical tests showed that these differences were negligible. 

\section{Conclusion}
\label{sec:conclusion}

In this paper, we proposed to utilize \vqw, a self-supervised speech representation learned without labeled data, to achieve A2O seq2seq VC. 
Using the same Transformer architecture, we found from the objective and subjective evaluation results that the proposed A2O method can outperform an O2O baseline system trained with parallel data. Moreover, our method is robust against limited training data and can be trained with only 5 min of data.
In the future, we plan to investigate self-supervised speech representations learned using different algorithms and objectives to determine what information is crucial for VC.
Also, as demonstrated in \cite{xlsr}, the ability of self-supervised speech representations to generalize to multiple or even unseen languages opens the possibility of applying our proposed method to cross-lingual VC, which will be another important future direction.

\section{Acknowledgements}
\vspace*{-1.5mm}
This work was partly supported by JST, CREST Grant Number JPMJCR19A3, and JSPS KAKENHI Grant Number 17H06101.

\bibliographystyle{IEEEbib}
\bibliography{ref}

\end{document}